\documentclass[reprint, prl, showkeys]{revtex4-2}

\usepackage{hyperref}

\usepackage[utf8]{inputenc}
\usepackage[english]{babel}
\usepackage{amsmath}
\usepackage{latexsym}
\usepackage{graphicx}
\usepackage{float}
\usepackage{multirow}
\usepackage{amssymb}
\usepackage{mathrsfs}
\usepackage{lineno}
\usepackage{lipsum}

\begin{document}

\title{Magnetization of $\mathrm{Zn_{1-\textit{x}}Co_{\textit{x}}O}$ nanoparticles: single-ion anisotropy and spin clustering.}

\author{X. Gratens$^{a}$, B. de Abreu Silva$^{a}$, M. I. B. Bernardi$^{b}$,  H. B. de Carvalho$^{c}$, A. Franco Jr$^{d}$ and  V.A. Chitta$^{a}$}
\affiliation{$^{a}$Instituto de Física, Universidade de São Paulo, 05315-970 São Paulo, SP, Brazil}
\affiliation{$^{b}$Instituto de Física de São Carlos, Universidade de São Paulo, 13560-970 São Carlos, SP, Brazil}
\affiliation{$^{c}$Instituto de Ciências Exatas, Universidade Federal de Alfenas – UNIFAL-MG, 37133-840 Alfenas, MG, Brazil}
\affiliation{$^{d}$Instituto de Física, Universidade Federal de Goiás, 74690-900 Goiánia, GO, Brazil}
\begin{abstract}
The magnetization of $\mathrm{Zn_{1-\textit{x}}Co_{\textit{x}}O}$ (0.0055 $\leq$ \textit{x} $\leq$ 0.073) nanoparticles has been measured as a function of temperature $T$ (1.7 K $\leq$ $T$ $\leq$ 10 K) and for magnetic field up to 65 kOe using a SQUID magnetometer. Samples were synthesized by three different growth methods: microwave-assisted hydrothermal, combustion reaction and sol-gel. For all studied samples, the magnetic properties derive from the antiferromagnetic (AF) spin clustering due to the Co$^{2+}$ nearest neighbors. At \textit{T} $\geq$ 6 K, the magnetization of the Co$^{2+}$ ions has a Brillouin-type behavior, but below 6 K, it shows a notable deviation. We have shown that the observed deviation may be derived from single-ion anisotropy (SIA) with uniaxial symmetry. Results of fits show that the axial-SIA parameter \textit{D} (typically \textit{D} = 4.4 K) is slightly larger that the bulk value \textit{D} = 3.97 K. No significant change of \textit{D} has been observed as a function of the Co concentration or the growth process. For each sample, the SIA fit gave also the effective concentration ($\overline{x}$) corresponding to the technical saturation value of the magnetization. Comparison of the concentration dependence of $\overline{x}$ with predictions based on cluster models shows an enhancement of the AF spin clustering independent of the growth method. This is ascribed to a clamped non-random distribution of the cobalt ions in the nanoparticles. The approach of the local concentration ($x_{L}$) has been used to quantify the observed deviation from randomicity. Assuming a ZnO core/ $\mathrm{Zn_{1-\textit{x}}Co_{\textit{x}}O}$ shell nanoparticle, the thickness of the shell has been determined from the ratio $x_{L}$/$x$.           
    
\end{abstract}

\maketitle

\section{I. Introduction}

Diluted magnetic semiconductors (DMSs) based on transition metal (TM) doped ZnO material have attracted great attention in the last two decades due to the possibility to induce a ferromagnetism at room temperature via carrier concentration manipulation \cite{Dietl}. Despite a large number of studies, the results especially concerning the nature of the observed ferromagnetism in TM doped ZnO are still questionable. On the other hand, a definite conclusion about the \textit{intrinsic} antiferromagnetic (AF) properties of ZnO doped with Mn$^{2+}$ or  Co$^{2+}$ ions without extra carriers is much easy to draw \cite{GratensZnO,Ambrosio,Sati,Yoon}. It was found that both $\mathrm{Zn_{1-\textit{x}}Mn_{\textit{x}}O}$ \cite{GratensZnO} and $\mathrm{Zn_{1-\textit{x}}Co_{\textit{x}}O}$ \cite{Ambrosio,Sati,Yoon} in bulk phase including polycrystal powders are typical members of the well known $\mathrm{A^{II}_{1-\textit{x}}Mn^{ }_{\textit{x}}B^{VI}}$ (A = Cd, Zn; B = Te, Se, S) \cite{Shapira1, Foner, Shapira2, Shapira3} DMS family characterized in the dilute limit by a paramagnetic behavior and an AF spin clustering due to strong exchange interaction between nearest neighbors magnetic ions. The magnetization is then characterized by an apparent saturation at low magnetic field and by the existence of magnetization steps (MSTs) at high field \cite{Shapira4}. The magnetic susceptibility ($\chi$) also affected by the AF spin clustering exhibits a Curie Weiss behavior in the high temperature regime and a deviation from the Curie Weiss Law in the low temperature region in the form of a downturn in the 1/$\chi$ vs temperature graph \cite{Yoon, Shapira1}.   

One of the most effective technique to study the spin clustering effect in DMSs is the MST method \cite{Shapira4}. This technique measures the exchange interaction constants, the anisotropic parameters and gives the relative population of the different types of spin clusters. One usuful "\textit{tool}" directly derived from the MST method is the technique of the apparent (or technical) saturation value of the magnetization ($M_{S}$). Here the technique yields to the population of all the clusters with a ground state ($S_{T}$) different to zero and to the effective magnetic-ion concentration $\overline{x}$ related to $M_{S}$. \cite{Shapira5}. 
Comparison of experimental value of $\overline{x}$ with predictions using clusters models \cite{Shapira4} gives relevant informations about the type of the dominant exchange interaction and the distribution of the magnetic ions over all the cation sites. It was shown for example that the Mn distribution in $\mathrm{A^{II}_{1-\textit{x}}Mn^{ }_{\textit{x}}B^{VI}}$ bulk samples is random \cite{Shapira5}.

Low dimensional DMSs nanostructures  with two, one and zero dimensional shapes have attracted growing research interest because of the combination of both quantum confinement and magnetic size effects. 
One of the main expected change of the magnetic features for the DMSs nanostructures is the reduction of the AF spin clustering and consequently an enhancement of the paramagnetism. The size of this effect is connected to the magnetic-ion distribution ranging from three to two, one or zero dimensional spin distribution as a function of the surface/volume ratio of the nanostructure as shown for digitized layers \cite{Crooker}. On the other hand, in the case of nanostructures prepared by solution synthesis methods, the well know difficulty of the dopant incorporation inside the nanostructures \cite{Erwin,Acharva} may produce non-random distribution of the magnetic ions and consequently an enhancement of the spin clustering instead of the paramagnetism. 
 
In this paper, we report the results of the spin clustering investigation performed on bulk like spherical shaped $\mathrm{Zn_{1-\textit{x}}Co_{\textit{x}}O}$ nanoparticles (NPs) with a typical diameter of 25 nm prepared by three growth process. Thirteen samples with the Co concentration ranging from 0.0055 to 0.073 have been investigated. The concentration of the Co$^{2+}$ ions has been determined from Curie-Weiss fit of the $\chi$ vs temperature data.
Usually, in DMSs, the magnetization curve at low temperature can be fitted to the Brillouin function. However, in the studied $\mathrm{Zn_{1-\textit{x}}Co_{\textit{x}}O}$ nanoparticles, we observed that the magnetization departs from the Brillouin behavior at low temperature. Similar deviation has been pointed out in previous work for polycrystalline powder samples \cite{Yoon}, but its clear interpretation was lacking to date. Here, we have successfully reproduce the observed deviation by considering the huge single-ion anisotropy (SIA) with uniaxial symmetry reported for bulk crystal \cite{Sati} in the calculation of the nanoparticles magnetization. The axial-SIA parameter $D$ of the nanoparticles was also determined. Finally, by using the apparent saturation method, conclusion on the spatial distribution of the magnetic ions into the nanoparticles can be drawn.

\section{II.Experiment}
The studied $\mathrm{Zn_{1-\textit{x}}Co_{\textit{x}}O}$ NPs were produced by three different growth methods: the microwave-assisted hydrothermal (MAH) \cite{Rafael}, combustion reaction (CR) \cite{Franco} and sol-gel (SG) \cite{Spanhel}. Systematic structural analysis of the samples from MAH and CR batches have been already published \cite{Rafael,Franco}. The results confirm the dilution of the Co ions into the cationic sub-lattice. Secondary phases of cobalt oxide ($\mathrm{Co_{3}O_{4}}$, CoO), Co-rich phase ($\mathrm{Co_{0.8}Zn_{0.2}O}$ \cite{Mesquita}) or metallic Co were not detected. A good phase purity is also found in the SG samples \cite{bruno}. The studied nanoparticles are spherical in shape with an average diameter of about 20 nm for MAH and CR samples \cite{Rafael,Franco} and around 30 nm for the SG nanoparticles \cite{bruno}.

The magnetic measurements were performed using a superconducting quantum interference device (SQUID) magnetometer. The magnetic susceptibility was obtained by measuring the magnetization (\textit{M}) for low magnetic fields (\textit{H}) and by using $\chi$ = $M$/$H$. For three samples a small ferromagnetic contribution is observed at 300 K in addition to the large paramagnetic response of the Co$^{2+}$ ions. In this situation, $\chi$ has been determined from the slope ($\chi = \Delta M $/$\Delta H $) of the linear variation of \textit{M} vs \textit{H} traces well above the saturation of the FM contribution. For all samples the susceptibility vs temperature trace shows no signal from impurity phase.  The Co$^{2+}$ concentration \textit{x} has been determined by fitting the magnetic susceptibility ($\chi$) between 200 and 300 K to a sum of a Curie-Weiss susceptibility and a constant $\chi _{d} $ representing the lattice diamagnetism. The spin \textit{S} = 3/2 and the isotropic Landé factor \textit{g} = 2.263 \cite{Sati} were used in the Curie-Weiss expression. 
The magnetization has been measured as a function of $H$ and for temperature $T$ in the range 1.7 K $\leq$  \textit{T} $ \leq$ 8 K.  The data have been taken with \textit{H} up to 65 kOe, and the magnetic reversibility has been confirmed by hysteresis loop measurements.

\section{III. Magnetization results and discussion}

\begin{figure}
\centering
\includegraphics[width=8.00 cm]{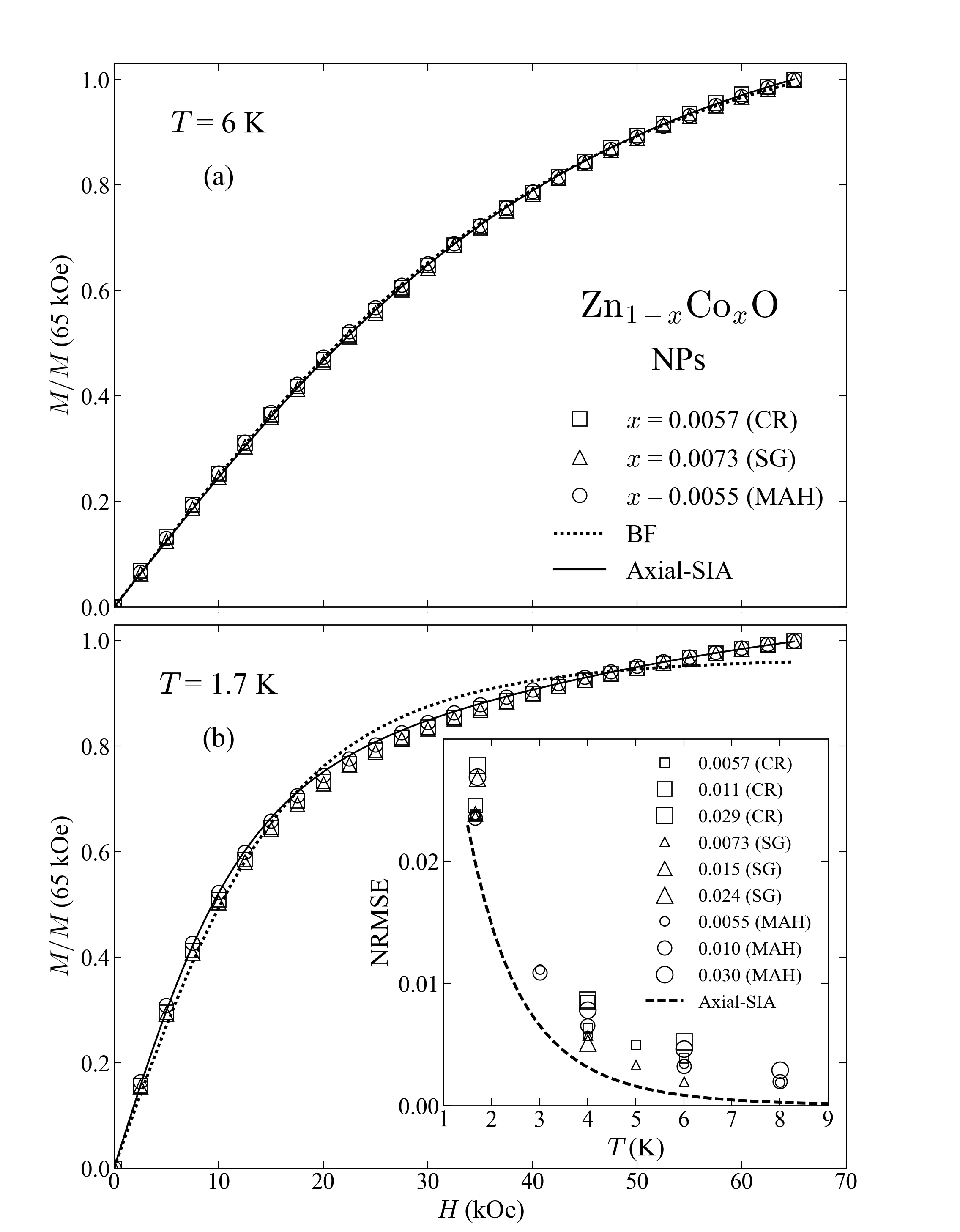}
\caption{Magnetization curves of samples with \textit{x} = 0.0057 (CR), 0.0073 (SG) and 0.0055 (MAH) at \textit{T} = 6 K (a) and \textit{T} = 1.7 K (b). $M$ has been normalized to its value at $H$ = 65 kOe. Symbols represent the experimental data. Dotted lines are the Brillouin function (BF) fitted curves obtained for \textit{x} = 0.0057 (CR). Solid lines represent the fitted curves obtained by using the axial-SIA model for the same sample. Inset of Figure 1 (b): Normalized root mean square error (NRMSE) of the BF fits as a function of $T$ for samples with different concentrations. Dashed line represents the $T$ - dependence of the NRMSE values obtained by fitting the calculated $M$ curves for the nanoparticles using the Brillouin function. The calculations were performed by using the axial-SIA model and the bulk value for \textit{D}.}
\label{Figure1}
\end{figure}

Figure 1 shows the magnetization curves measured at temperature \textit{T} = 6 K (Fig. 1(a)) and \textit{T} = 1.7 K (Fig. 1(b)) for the samples with the lowest concentration of each growth technique, $x$ = 0.0055 (MAH), $x$ = 0.0073 (SG) and $x$ = 0.0057 (CR). Data include the diamagnetic correction from the ZnO lattice and \textit{M} was normalized to its value measured at \textit{H} = 65 kOe. For both temperatures, the magnetization traces of the three samples show similar shape. Figure 1 displays also the curves obtained by fitting the experimental data of the sample $x$ = 0.0057 (CR) to the modified Brillouin function (BF) \cite{Gaj}. The parameters \textit{S} = 3/2 and  \textit{g} = 2.263 \cite{Sati} have been used in the BF expression. The fitting parameters were $\overline{x}$ and the effective temperature $\mathrm{\textit{T}_{eff}}$ which replaces the experimental temperature in order to take into account of the effect from the distant neighbors. For \textit{T} = 6 K, the fitted curve agrees quite well with the experimental data. However, at \textit{T} = 1.7 K, the BF gives a poor fit of the magnetization curve and clearly underestimates the value of $\overline{x}$. The same feature is also observed for the other samples with higher concentrations as shown in Fig. 2. We may observed that the departure of the magnetization curve from the BF behavior is almost identical for all samples, independently of the sample production method. This is also confirmed by the inset of Fig. 1, where the normalized root mean square error (NRMSE) values calculated between the BF fitted curves and the experimental data are plotted as a function of the temperature for nine samples with different $x$. The NRMSE values have been obtained by normalizing the error values to the maximum value of $M$. Lower values of NRMSE would indicate a better performance of the fit. For each temperature, we obtained comparable values of NRMSE.  The figure shows also a monotonically increase of NRMSE as the temperature decreases.

\begin{figure}
\centering
\includegraphics[width=8.00 cm]{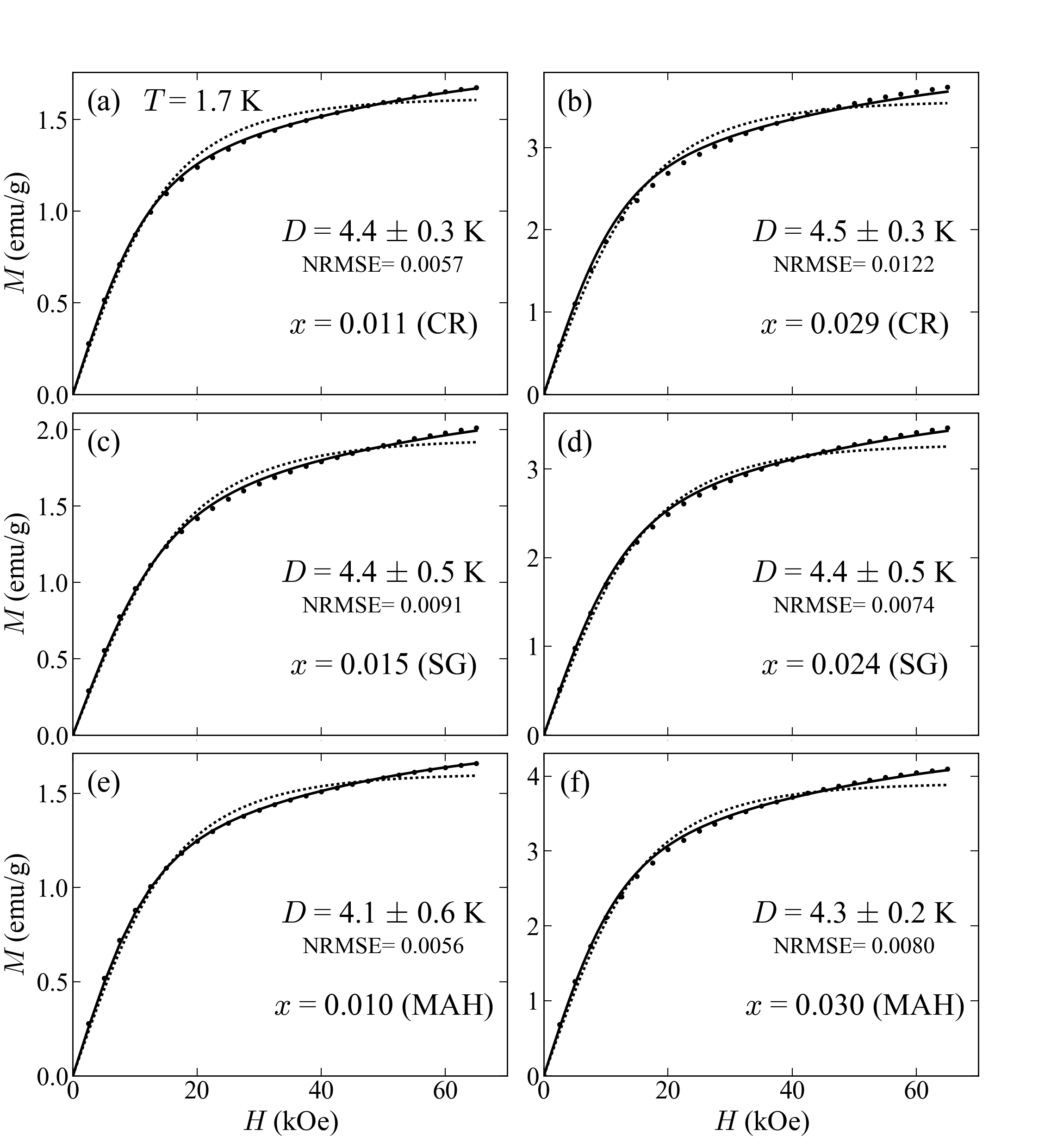}
\caption{Magnetization curves at \textit{T} = 1.7 K for samples with \textit{x} = 0.011 (a) and \textit{x} = 0.029 (b) grown by combustion reaction method, with \textit{x} = 0.015 (c) and \textit{x} = 0.024 (d) obtained from sol-gel process, and with \textit{x} = 0.010 (e) and \textit{x} = 0.030 (f) grown using microwave-assisted hydrothermal route. The experimental data are represented by dot symbols. Dotted lines are the BF fitted curves. Solid lines were obtained by fitting the experimental data to the axial-SIA model.}
\label{Figure1}
\end{figure}

As described in previous works \cite{Ambrosio,Sati}, the magnetization of $\mathrm{Co^{2+}}$ ions in bulk $\mathrm{Zn_{1-x}Co_{x}O}$ samples displays a strong axial-SIA with the axis symmetry along the \textbf{c} axis of the wurtzite structure. The magnetic behavior of $\mathrm{Co^{2+}}$ ions can then be described by using an effective spin \textit{S} = 3/2  and the conventional spin Hamiltonian \cite{Abragam}:   

\begin{equation} \label{eq1}
\begin{split}
\mathscr{H} = g_{\parallel}\mu_{B}H_{z}S_{z} + g_{\perp}\mu_{B}(H_{x}S_{x}+H_{y}S_{y})\\
+ D \Big[S_{z}^{2} - \dfrac{1}{3}S(S+1)\Big]
\end{split}
\end{equation}

The quantifization \textit{z} axis is taken along the \textbf{c} axis of the wurtzite lattice. $H_{x}$, $H_{y}$ and $H_{z}$ are defined using $\theta$ and $\varphi$ the polar and azimuthal angles of \textbf{H}. $g_{\parallel}$ and $g_{\perp}$ are the effective \textit{g} factors parallel and perpendicular to \textit{z}. Figure 3 (a) highlights the resulting anisotropy of the magnetization curves for $\mathrm{Zn_{1-x}Co_{x}O}$ bulk. Here, the traces have been calculated at $T$ = 1.7 K for \textbf{H} parallel and perpendicular to \textbf{c} by diagonalization of the spin Hamiltonian matrix of Eq. (1) using the bulk parameters $ g_{\parallel} = 2.236, g_{\perp}= 2.277$ and $D/k_{B}=3.971$ K  \cite{Sati}. Because the magnetization with \textbf{H}  $\perp$ \textbf{c} ($\mathrm{M_{\perp}}$) rises faster than $\mathrm{M_{\parallel}}$ for \textbf{H} $\parallel$ \textbf{c}, the anisotropy is \textit{easy plane} like type. The magnetizations curved of Fig. 3 (a) are very similar to both theoretical and experiemental traces presented in Ref \cite{Sati} for epitaxial layer.

Quite naturally the next step was to applied the SIA model to calculate the magnetization of the nanoparticles based on the assumption of a random orientation of them with respect to the magnetic field. Following this approach, the magnetization of the nanoparticles is the summation of the magnetizations corresponding to every  orientation \{$\theta$, $\varphi$\} of \textbf{H}. Here, the orientations have been represented by a rectangular grid with fixed increments of $\theta$ and $\varphi$\ on the unit sphere. Because of uniaxial symmetry, the magnetization depends on the angle $\theta$ only, and the interval of calculation can be reduced to the thin slice of the unit sphere defined by $\Delta \varphi$ = 1$^{\circ}$ and $0 \leq \theta \leq \dfrac{\pi}{2}$.  

The magnetization of the $\mathrm{Zn_{1-x}Co_{x}O}$ nanoparticles may then be numerically computed using:

\begin{equation} \label{eq1}
\textit{M} = \sum\limits_{\theta =0}^{90} M (\theta)sin(\theta) \Delta \theta   
\end{equation}

\begin{figure}
\centering
\includegraphics[width=8.00 cm]{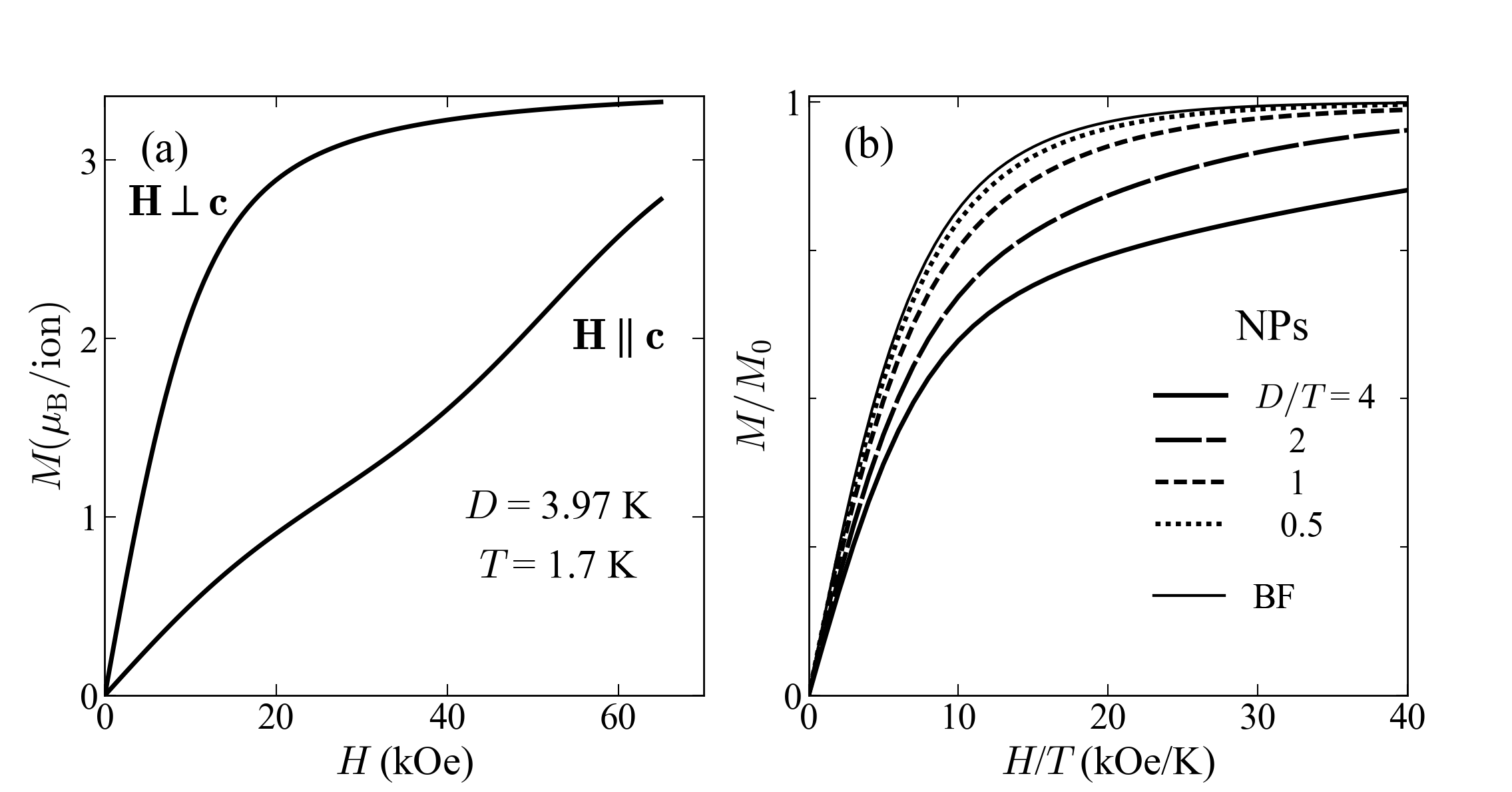}
\caption{(a) Simulations of the magnetization curves for $\mathrm{Zn_{1-\textit{x}}Co_{\textit{x}}O}$ bulk based on the axial-SIA model. The traces were computed for  \textit{T} = 1.7 K  and \textbf{H} parallel and perpendicular to the \textbf{c} axis using the bulk parameters \cite{Sati}.(b) Theoretical \textit{M} vs $H$/$T$ curves based on the axial-SIA model assuming randomly oriented nanoparticles calculated for different values of $D$/$T$. \textit{M} is normalized to its true saturation value $M_{0}$. For comparison, the Brillouin function is displayed in the figure.}
\label{Figure1}
\end{figure}

where $\Delta \theta$ is the increment angle value used in the summation. Simulations based on Eq.(2) have shown that magnetization curves calculated for different values of $\Delta \theta$ with $\Delta \theta \leq 0.1 ^{\circ} $ are almost identical.

Figure 3 (b) shows the $M$ vs $H/T$ curves computed using Eq. (2) for different values of $D$/$T$. The $g_{\parallel}$ and $g_{\perp}$ values of the bulk have been used here. The magnetization is normalized to its true saturation value $M_{0}$. The figure shows the evolution with increasing $D$/$T$ ratio from the Brillouin function, to a non-Brillouin shaped magnetization curve characterized by a high field saturation. The curve can then be roughly described by a fast rise of $M$ at low fields which is the manifestation of the contributions with $\theta$ near 90 deg ($\mathrm{M_{\perp}}$), followed by a ramp due to the contributions with $\theta$ near 0 deg ($\mathrm{M_{\parallel}}$).

To illustrate the change in the curve shape, we have performed modified BF fits of the magnetization data calculated using the bulk $D$ value, in the 0 - 65 kOe field range and for different values of $T$. The corresponding values of NRMSE are displayed as a function of the temperature in the inset of Fig. 1(b). Clearly, the SIA model reproduces quite well the temperature dependence of the experimental NRMSE values.  

  Finally, least-square fits of the experimental magnetization curves measured at $T$ = 1.7 K have been performed by using Eq. (2) and the Nelder-Mead simplex method \cite{Nelder} with $\overline{x}$ and $D$ as fitting parameters. Here, the experimental temperature $T$ = 1.7 K has been used as input in the calculation of the magnetization. The curves obtained from the SIA model fits are displayed in Fig. 1 for sample $x$ = 0.0057 (CR) and in Fig. 2 for the others six samples. Both figures show an excellent agreement between the SIA model fitting and the experimental data, in contrast with the BF fits. The gain in the fitting accuracy is also confirmed quantitatively by much more lower values of NRMSE (Table 1 and Fig. 2) than the typical value of about 0.025 given by the BF fits. Based on the fitting results at $T$ = 1.7 K the agreement between the curve predicted by the SIA model and the experimental data taken at different temperature has been examined. In all cases, the agreement was very satisfactory. An example is given in Fig. 1 for sample $x$ = 0.0057 (CR) at $T$ = 6 K.

The overall fitting results point out that the axial-SIA parameter $D$ is independent (within the uncertainty) of both the Co concentration (at least for $x$ $\leq$ 0.03) and the growth method. Typically, we have $D$ = 4.4 K for the nanoparticles studied. This quoted value is slightly larger than the literature value of 4 K. However, due to the uncertainty in $D$ of about 10 \%, a clear evidence of $D$ change is difficult to draw for our samples. We may finally conclude that the single-ion anisotropy is bulk-like type in the studied nanoparticles. Comparison of the fitting results obtained from the two types of fit (SIA model and BF) shows a good agreement between the $\overline{x}$ values  (Table. 1) obtained from the SIA model fit and those determined by the BF fit of the experimental data at $T$ = 6 K. This is not surprising because the relative difference between the SIA model curve and the Brillouin function is only 0.1 \% for that temperature.

The analysis of the fitting results shows also a slight decrease of the fitting accuracy (see the NRMSE values in Fig. 1 and Fig. 2) with the increase in the Co  concentration. For samples with higher concentrations,  $x$ = 0.045 (CR), $x$ = 0.057 (MAH), $x$ = 0.0723 (MAH) and $x$ = 0.073 (CR), the SIA model gives a worst fit of the magnetization curves at $T$ = 1.7 K, but still better than the BF fits. This change may be ascribed, in addition to the effect of the distant neighbors, to the contribution of the AF clusters larger than the singles, whose population increases with increasing $x$. In fact, for Co$^{2+}$ ions which are in larger clusters, the local environment and the anisotropy may be different from the isolated ions. For these samples, the BF fits very well (as for the others samples) the experimental data at $T$ = 6 K, and the $\overline{x}$ values obtained for that temperature will be used in the next.

\begin{figure}
\centering
\includegraphics[width=8.00 cm]{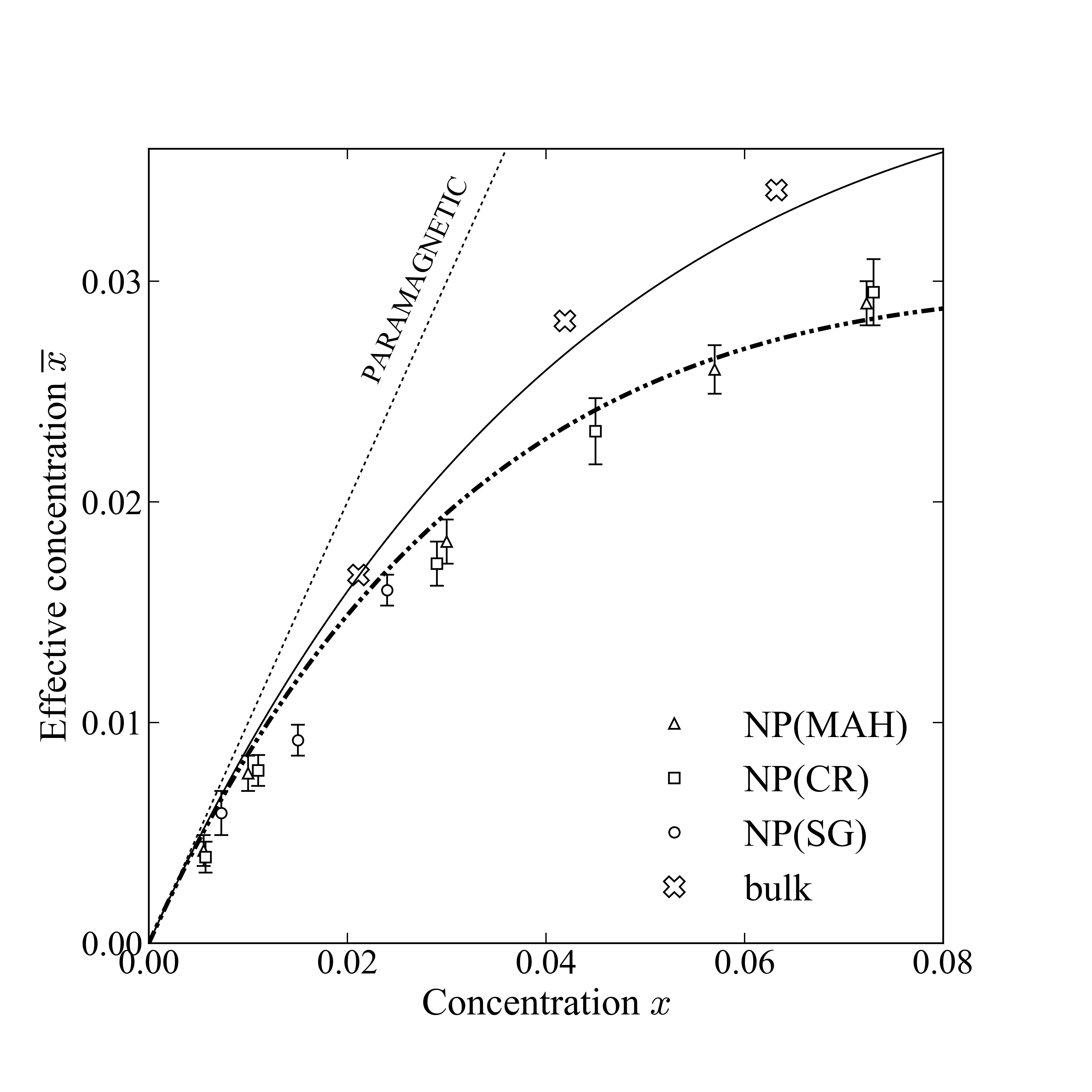}
\caption{Effective concentration as function of the Co concentration for $\mathrm{Zn_{1-x}Co_{x}O}$ nanoparticles. Data for polycrystal powder samples (Ref. \cite{Bouloudenine})  are also presented. The dotted line corresponds to the "\textit{pure}" paramagnetic behavior. The solid line is the calculated curve giving by Eq.(3), based on the $J_{1}$ cluster model and assuming a random distribution of the Co ions. The dashed-doted line is the fit of the NPs overall data to Eq. (3) and using the concept of local concentration ($x_{L}$). The fit gave $x_{L} \simeq 1.4 $ $x$. }
\label{Figure1}
\end{figure}

\begin{table*}[ht]
\centering
\renewcommand{\arraystretch}{1.5}
\begin{tabular}{|c|c|c|c|c|c|}\hline
Sample& $\overline{x}$ (BF) & NRMSE &$\overline{x}$ (SIA) & $D$ (K)& NRMSE  \\ \hline
0.0055 (MAH)     & 0.0041 & 0.0034  & 0.0043  & 4.4 $\pm$ 0.2 & 0.0047  \\ 
 \hline
0.0057 (CR)      & 0.0037  & 0.0039& 0.0039  & 4.4 $\pm$ 0.3 & 0.0067\\ \hline

0.0073 (SG)     & 0.0060   & 0.0019 & 0.0060 & 4.6 $\pm$ 0.3 & 0.0024 \\
 \hline
\end{tabular}
\caption{Comparison of the effective concentration $\overline{x}$ obtained from BF fit of the magnetization data at \textit{T} = 6 K and from axial-SIA fit of \textit{M} vs \textit{H} experimental curves at \textit{T} = 1.7 K. The uncertainly of  $\overline{x}$ is 0.0005.}
\label{Table1}
\end{table*}

Figure 4 displays the effective concentration determined for the studied NPs samples as a function of the Co concentration. For comparison, the "\textit{pure}" paramagnetic behavior $\overline{x}$ = $x$ (dotted line) and the data obtained for polycrystal powder samples Ref. \cite{Bouloudenine} are also displayed in the figure.

Predictions of $\overline{x}$  can be performed by using cluster models. In the simplest model the dominant exchange interaction is ascribed to the nearest neighbors pairs. This model often called $J_{1}$ model has been successfully used for II-VI DMSs with zinc blend lattice structure \cite{Foner}. For wurtzite type DMSs the $J_{1}$ model has been modified to take into account the symmetry unequivalence by symmetry of the nearest neighbors cations. In fact, each cation of the hcp lattice has two groups of unequivalent nearest neighbors cation sites, 6 nearest neighbors sites in the same plane perpendicular to the \textbf{c}-axis and other 6 sites out of the \textbf{c}-plane. The exchange constants $J_{1}^{in}$ and  $J_{1}^{out}$ associated to the two groups of nearest neighors ($in$ and $out$) are different. Both constants have been measured for $\mathrm{Cd_{1-\textit{x}}Mn_{\textit{x}}S(e)}$ \cite{Shapira6,Bindilatti}, $\mathrm{Zn_{1-\textit{x}}Mn_{\textit{x}}O}$ \cite{GratensZnO}, $\mathrm{Cd_{1-\textit{x}}Co_{\textit{x}}S(e)}$ \cite{Foner2} and in bulk $\mathrm{Zn_{1-\textit{x}}Co_{\textit{x}}O}$ \cite{Ambrosio}. For the last material, the average $\overline{J_{1}}$ = ($J_{1}^{in}$ + $J_{1}^{out}$)/2 is -21 K, with a huge 80 \% difference between the two $J_{1}$ values.

Based on cluster models, $\overline{x}$/$x$ can be expressed as the sum of the contributions of all clusters with ground state at zero magnetic field ($S_{T}(0)$) different from zero. The contribution of one of these clusters is related to its $S_{T}(0)$ value and population. For $\mathrm{Zn_{1-\textit{x}}Co_{\textit{x}}O}$, due to the large two  $J_{1}$ constants, the computational effort can be minimized by asserting $J_{1}^{in}$ = $J_{1}^{out}$ = $\overline{J_{1}}$ in the modified $J_{1}$ model for hcp lattice DMSs. In this work, $\overline{x}$/$x$ has been calculated for all the cluster types up to the quintets (cluster forming by five nearest neighbors magnetic ions) for $S$ = 3/2. The ratio $\overline{x}$/$x$ is then given by:

\begin{equation} \label{eq1}
\begin{split}
\overline{x}/x  = P_{S} + P_{OT}/3 + P_{CT}/9 + P_{PQ}/2 + P_{FQ}/6\\
+\sum \limits_{i=1}^{17} F\mathrm{_{V}}_{i} P\mathrm{_{V}}_{i} + P_{others}/7 
\end{split}
\end{equation}
where $P_{S}$, $P_{OT}$, $P_{CT}$, $P_{PQ}$ and $P_{FQ}$ are the probabilities that a magnetic ion belongs to singles, open triplets, closed triplets, propeller quartets and funnel quartets respectively \cite{Shapira4}. $P\mathrm{_{V}}_{i}$ is the probability of finding a spin in the i-th type of quintets.  The probability that a magnetic ion is in one type of cluster is derived from the cluster tables given in Ref.\cite{Valdir}.  
The arithmetic expressions of $P\mathrm{_{V}}_{i}$ as function of $x$ and the $F\mathrm{_{V}}_{i}$ factors (1/5, 3/5, 1/15 or 1/3) are given in Ref \cite{Gratens} for the seventeen types of quintets. The contribution of clusters larger than the quintets is included in the last term of Eq. (3)  by assuming that they are sextet string clusters with $S_{T}(0)$ = 1/7. $P_{others}$ is the probability of finding a magnetic ion in a cluster larger than the quintet. 

Figure 4 shows the theoretical curve of $\overline{x}$ as function of $x$ predicted by Eq. (3) assuming a random Co distribution. The comparison with the experimental data of bulk (polycrystal powder) samples shows a good agreement between the predicted and experimental traces, consistent with a random or nearly random Co distribution in the bulk material. On the other hand, large deviations from the random distribution can be observed in the figure for the nanoparticle samples. 
For these samples, the data are well below the predictions indicating that the actual number of clusters (different of the singles) are much larger than that calculated from a random distribution. The observed enhancement of the AF clustering effect may be assigned to a tendency of the magnetic ions to bunch together observed in other DMSs \cite{GratensSnEuTe}. The result is the existence of two different regions in the sample, one occupied and other avoided by the magnetic ions.

\begin{figure}
\centering
\includegraphics[width=6.00 cm]{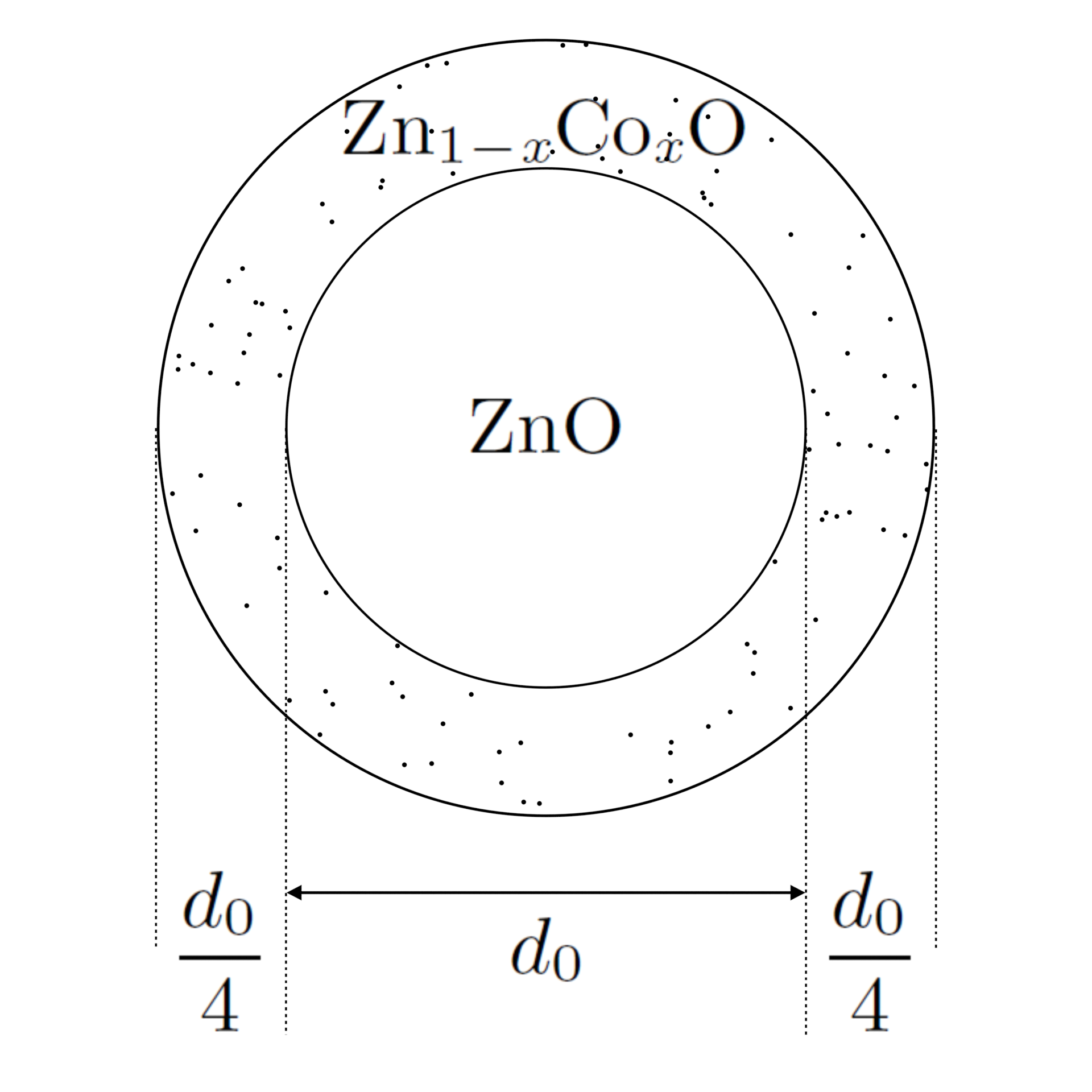}
\caption{Core/shell model of the nanoparticles.}
\label{Figure1}
\end{figure}

We may used the concept of the local concentration ($x_{L}$) \cite{GratensSnEuTe} to describe the nonrandom distribution of the Co ions. In this approach, $x_{L}$ which is the average concentration in the neighbourhood of the Co ions of the occupied regions replaces $x$ in Eq. (3).  The experimental data for the nanoparticles were fitted to Eq. (3) using $x_{L}$ as a fitting parameter. The fit (displayed in Fig. 4) gave $x_{L}$ = (1.4 $\pm$ 0.1)$x$.

According to previous works \cite{Erwin, Acharva, Rafael} on the difficulty to incorporate magnetic ions into the core of a nanoparticle, we have naturally associated the local concentration approach to the well know core/shell nanoparticles picture. In the simplest model, the nanoparticle is composed of a core of pure ZnO, surrounded by a shell of $\mathrm{Zn_{1-\textit{x}}Co_{\textit{x}}O}$ as shown in Fig. 5. In the shell, the Co distribution is assumed to be random with a concentration given by $x_{L}$. By neglecting any change of volume density of the two phases, the ratio of the core diameter  ($d_{0}$) to the NP diameter ($d$) is given by:           

\begin{equation} \label{eq1}
\begin{split}
\dfrac{d_{0}}{d} \approx \Big(~1-\dfrac{x}{x_{L}}\Big)^{1/3}
\end{split}
\end{equation}

Using the previous result $x_{L}$/$x$ = 1.4, we obtain that the thickness of the shell is about $d_{0}/4$.

We may also noted that fitting the data for samples with $x$ $\leq$ 0.03 gave a slight increase of the local concentration with $x_{L}$/$x$ = 1.8 $\pm$ 0.2, and consequently a decrease of the $\mathrm{Zn_{1-\textit{x}}Co_{\textit{x}}O}$ shell thickness. This result agrees with the improvement of the magnetic ions incorporation in the core of nanoparticles with decreasing NPs size \cite{Rafael}, but contradicts the results obtained for other DMS \cite{Petropoulos}.

\section{IV. Conclusion }
The present work brings to light from results of simple SQUID measurements two new and relevant features of $\mathrm{Zn_{1-\textit{x}}Co_{\textit{x}}O}$ nanoparticles. First, the observed deviation of the magnetization from the classical Brillouin behavior has been successfully explained by bulk-like axial-SIA. Secondly, we observe that the studied nanoparticles exhibit an enhancement of the AF clustering associated to  a clumped dispersion of the Co ions into the nanoparticle volume. This deviation from random distribution can be quantify by using the concept of local concentration and surprisingly is not depending on the growth technique. Based on the approach of a structured ZnO core / $\mathrm{Zn_{1-\textit{x}}Co_{\textit{x}}O}$ shell particle, the local concentration may be associated to the Co concentration in the $\mathrm{Zn_{1-\textit{x}}Co_{\textit{x}}O}$ shell. Finally, the ratio of the thickness of the shell to the diameter of the core can be derived from $x_{L}$/$x$. This work demonstrates the ability of the apparent saturation technique to probe the spacial Co distribution in the nanoparticle.

\begin{center}
$\mathbf(Acknowledgments)$
\end{center}
Support from CNPq (grant 306715/2018-0) and FAPESP (grant 2015/16191-5) is gratefully acknowledged.

\end{document}